\documentclass{aa}  
\usepackage{graphicx}
\usepackage{txfonts}
\begin{document}
   \title{The Pre-Main-Sequence Eclipsing Binary
	  ASAS~J052821+0338.5\thanks{Based on observations made with the Nordic
	  Optical Telescope}}

   \subtitle{}

   \author{H. C. Stempels\inst{1}\fnmsep\thanks{email: \mbox{Eric.Stempels@st-andrews.ac.uk}}
           \and
	   L. Hebb\inst{1}
	   \and
	   K. G. Stassun\inst{2}
	   \and
	   J. Holtzman\inst{3}
	   \and
	   N. Dunstone\inst{1}
	   \and
	   L. Glowienka\inst{4,5}
	   \and
	   S. Frandsen\inst{5}
          }

   \offprints{H. C. Stempels}

   \institute{School of Physics \& Astronomy, University of St Andrews, North
              Haugh, St Andrews KY16 9SS, Scotland
              \and
	      Physics and Astronomy Department, Vanderbilt University,
	      Nashville, TN 37235, USA
	      \and
	      Astronomy Department, New Mexico State University, Las Cruces, NM 88003, USA
	      \and
	      Nordic Optical Telescope, Apartado 474, 38700~Santa Cruz de La Palma, Spain
	      \and
	      Department of Physics and Astronomy, {\AA}rhus University, Denmark
		                  }

   \date{Received; accepted}

 
  \abstract
   {}
   {In this paper we present the first results of a detailed spectroscopic and
   photometric analysis of the $V = 11.7^{\rm m}$ eclipsing binary
   \object{ASAS~J052821+0338.5}.
   }
   {With the FIES spectrograph at the Nordic Optical Telescope we have obtained
   a series of high-resolution spectra \mbox{($R \approx 47000$)} covering the
   entire orbit of the system. In addition we obtained simultaneous broadband
   photometry from three small aperture telescopes. From these spectroscopic and
   photometric data we have derived the system's orbital parameters and
   determined the fundamental stellar parameters of the two components.
   }
   {Our results indicate that ASAS~J052821+0338.5 is a K1/K3 pre-main-sequence
   eclipsing binary, with component masses of $1.38\,{\rm M}_{\sun}$ and
   $1.33\,{\rm M}_{\sun}$ and a period of 3.87 days, located at a distance of
   $280 \pm 30$ pc. The kinematics, physical location and the evolutionary
   status of the two stars suggest that ASAS~J052821+0338.5 is a member of the
   $\sim 11$ Myr old Orion OB1a subassociation. The systems also exhibits smooth
   $\sim 0.15^{\rm m}$ out-of-eclipse variations that are similar to those found
   in RS~CVn binaries. Furthermore the parameters we derived are consistent with
   the $10$--$13$ Myr isochrones of the popular Baraffe stellar evolutionary
   models.
   }
   {}

   \keywords{stars: pre-main sequence -- stars: binaries: eclipsing --
             stars: binaries: spectroscopic -- stars: fundamental parameters --
	     stars: individual: ASAS~J052821+0338.5
	     }

   \maketitle
%

\section{Introduction}

The advantageous geometry of detached double-line eclipsing binary systems makes
it possible to measure stellar radii and masses to very high accuracy (1-2\%,
Andersen \cite{andersen91}). Such accurate parameters are of crucial importance
in providing observational constraints on stellar evolutionary models. However,
very few of the known double-lined eclipsing binaries are stars at early stages
of evolution, leaving the models in this domain with only few empirical points
of reference.

The list of known double-lined eclipsing binaries with components that are in
the pre-main-sequence (PMS) phase of evolution is not long. Up to now, only five
low-mass double-lined eclipsing binaries in which both components are PMS
objects have been reported, RXJ~0529.4+0041A (Covino et al. \cite{covino00},
Covino et al. \cite{covino04}), V1174~Ori (Stassun et al. \cite{stassun04}), the
brown dwarf eclipsing binary 2MASS~J05352184-0546085 (Stassun et al.
\cite{stassun06, stassun07}) and, very recently, JW 380 (Irwin et al.
\cite{irwin07b}) and Par 1802 (Cargile et al. \cite{cargile07}). In addition,
EK~Cep and TY~CrA (Popper \cite{popper80}; Popper \cite{popper87}; Andersen
\cite{andersen91}; Casey \cite{casey98}) are known to harbour one PMS component
and one main-sequence component. 

In this paper, we announce the discovery that \mbox{ASAS J052821+0338.5} (also
known as \mbox{GSC2.2 N300311274}, \mbox{2MASS 05282082+0338327} and
\mbox{USNO-B1.0~0936-0073790}) is a PMS eclipsing binary system. In addition,
the system could be associated with the ROSAT X-ray source \mbox{1RXS
J052820.4+033823}, located only $11^{\prime\prime}$ away, within ROSAT's
positional error margin of $14^{\prime\prime}$. Using new high-cadence
photometric and high-resolution spectroscopic observations we have determined
the system's fundamental properties and confirm the PMS nature of both
components.

\section{Observations}

\subsection{Photometry}

ASAS~J052821+0338.5 was first identified in the All-Sky Automated Survey (ASAS,
Pojmanski \cite{pojmanski02}) as a detached eclipsing binary system with a
maximum, out-of-eclipse $V$-band magnitude of $V = 11.7^{\rm m}$ and a period of
$P = 1.9364$~days. In addition, the target is listed in several large
photometric databases consolidated in the Naval Observatory Merged Astronomical
Dataset (NOMAD-1.0, Zacharias et al. \cite{nomad}), which provide optical and
infra-red magnitudes of $B = 12.68^{\rm m}$, $J = 9.636^{\rm m}$, $H =
9.098^{\rm m}$ and $K = 8.961^{\rm m}$.

In the ASAS survey, 171 $V$-band measurements of the object were obtained during
the period February 2001--April 2006 with a median sampling rate of
0.25~d$^{-1}$. The resulting lightcurve exhibits periodic eclipses with a depth
of $\sim 0.3^{\rm m}$ and scatter in the out-of-eclipse phases with an rms
amplitude of $0.15^{\rm m}$, likely due to an evolving spot configuration. We
supplemented the long-term photometry from ASAS with additional precise,
high-cadence observations obtained with several small aperture telescopes.

\begin{figure}
\centering
\includegraphics[angle=0,width=0.95\columnwidth]{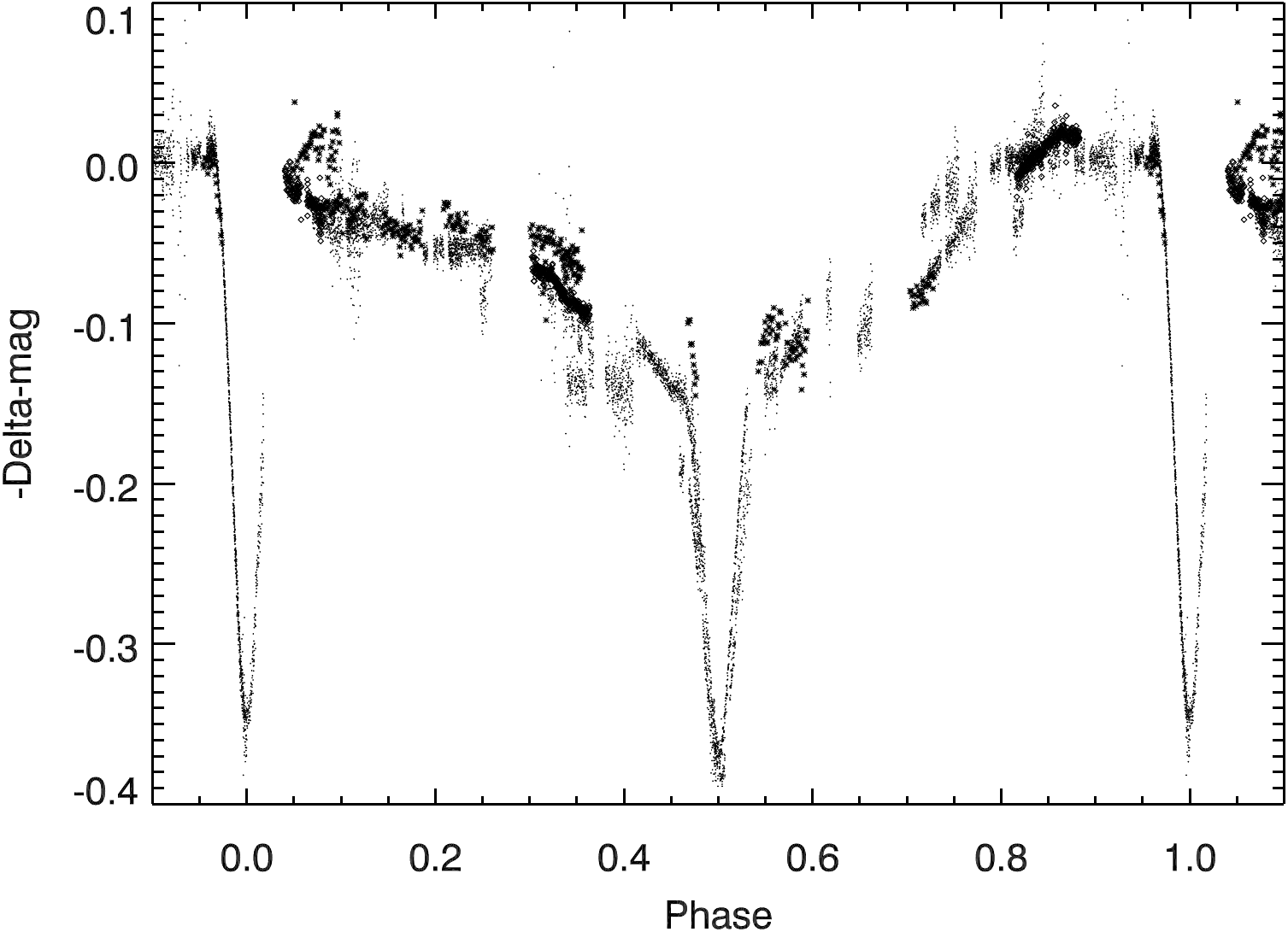}
\caption{All differential photometry of ASAS~J052821+0338.5 collected by the
NMSU-1m, JGT-1m and 3MT, phase-folded with the orbital period of $3.8729$~days
and epoch, $T_0 = 2454110.8500$.
The error bars on the data are smaller than the plotted symbols, and estimates
of the errors are given in the text.}
\label{fig:folded}
\end{figure}

\subsubsection{High cadence, high precision photometry}
\label{sec:phot}

The New Mexico State University 1-m telescope (NMSU-1m) provided the majority of
the follow-up photometry. A total of 8805 $V$-band measurements of the target
were acquired over 31~nights between January~10--April~17, 2007. The data were
processed with standard data reduction procedures including 2D bias and overscan
subtraction, flat-fielding, and aperture photometry. To derive the differential
photometry lightcurve, the magnitudes of three nearby, bright stars were average
combined into a non-variable reference object which was subtracted from the
instrumental magnitude of the target star. On February 4, 5, and 7, three
additional nights of $V$-band photometry were obtained with the 1-m James
Gregory Telescope (JGT-1m, \mbox{St. Andrews,} UK). These data can be directly
compared with the NMSU-1m data as the same basic reduction procedures were
applied to the images and the same three reference stars were used to derive the
differential photometry. The $V$-band differential photometry is normalized so
that the magnitude of the phase between 0.94--0.96 (before the primary eclipse)
has a median value of zero.

\begin{figure}
\centering
\includegraphics[angle=0,width=\columnwidth]{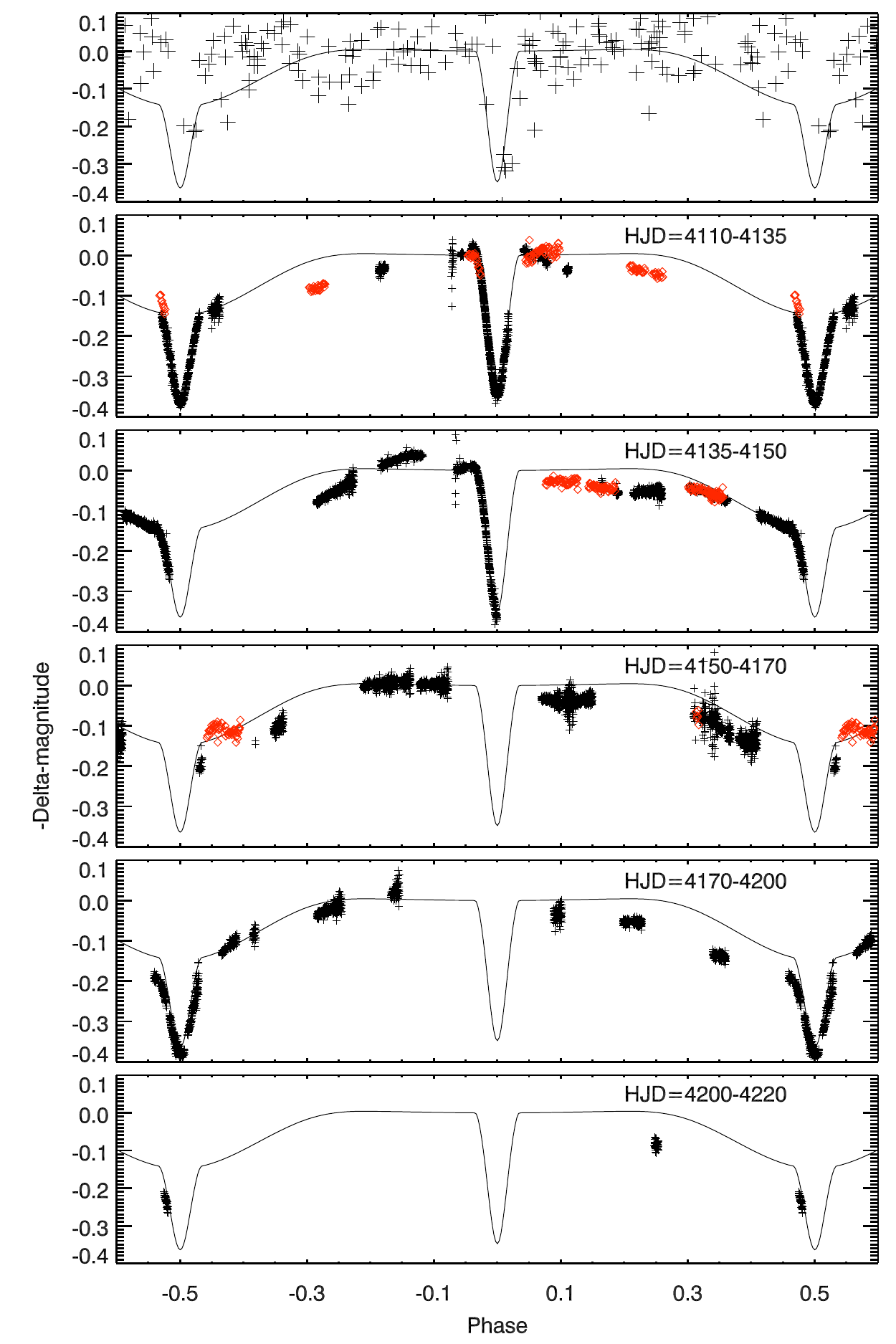}
\caption{This figure illustrates the evolution of out-of-eclipse variability on
ASAS~J052821+0338.5 during the four months (January--April, 2007) following the
spectroscopic observations. The original ASAS photometry is shown in the top
panel. Subsequent panels show the $V$-band lightcurve obtained with the NMSU-1m
and JGT-1m telescopes (black crosses) as well as the red (RG630) lightcurve
obtained with the 3MT (grey diamonds). The data  are broken into 2--4~week
intervals. The time span (in ${\rm HJD} - 2450000$) covered by the data in each
panel is given in the top right hand corner. The solid line is a simple one-spot
model, plotted as a reference. }
\label{fig:lcpanel}
\end{figure}

Time-series photometry in a redder waveband was obtained with the 0.5m 3-Mirror
Telescope (3MT; Cambridge, UK) from January--March, 2007. The filter (RG630)
plus the CCD efficiency curve for the detector selects a broadband wavelength
region which is between the Johnson $R$ and Cousins $I$-band filters. These data
were reduced in a standard way with a version of the Cambridge Astronomical
Survey Unit (CASU) data reduction and photometry pipeline (Irwin \& Lewis
\cite{irwin01}; Irwin et al. \cite{irwin07a}) modified for the 3MT. These data
are normalized to match the $V$-band photometry, so that the phases before the
primary eclipse have a median value of zero.

\begin{figure*}
\centering
\includegraphics[angle=90,width=0.95\textwidth]{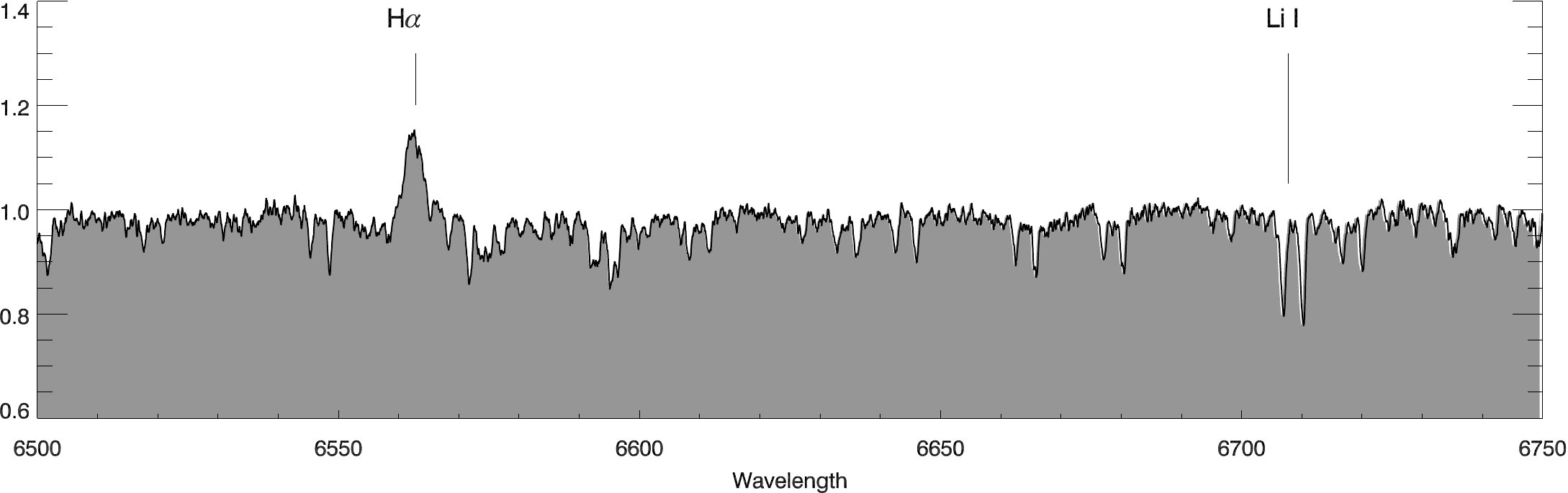}
\caption{This spectrum of ASAS~J052821+0338.5 was obtained at phase 0.64, and
shows clearly the double-lined nature of the spectrum, the emission in H$\alpha$
as well as the strong absorption in the Li {\sc i} 6708 line. }
\label{fig:pmsfeat}
\end{figure*}

The final differential photometry lightcurve of all telescopes folded with the
ephemeris derived below is shown in Fig.~\ref{fig:folded}. In
Fig.~\ref{fig:lcpanel} we show the original ASAS photometry (top panel) as well
as our small-telescope data, divided into panels with photometry obtained within
consecutive 2--4 week time intervals, the typical timescale for a stable
starspot configuration. The ASAS photometry has a precision of $\sim 0.05^{\rm
m}$. The small-telescope $V$-band photometry (crosses) has a sampling of $<
1$~minute, and has an uncertainty ranging from $\sim 0.003$ to $0.010^{\rm m}$,
mainly due to variable systematic errors such as changing sky transparency. We
opted to exclude 17 measurements which had unusually large photometric errors of
$> 0.05$ mag. The red wavelength data from the 3MT (diamonds) has a precision of
$0.006$--$0.008^{\rm m}$ and has a slightly higher scatter and lower sampling
due to the smaller aperture telescope used.

The lightcurve shows deep primary and secondary eclipses ($\sim 0.37^{\rm m}$),
a period ($P \approx 3.87$~days) twice what was determined from the
ASAS~photometry alone, and large-amplitude out-of-eclipse variability ($\sim
0.15^{\rm m}$). The out-of-eclipse variations are present for the duration of
our four month observing campaign, but they are not completely stable. There are
small variations in the lightcurve and in the depth of the secondary eclipse
(approximately $\pm 0.01^{\rm m}$), probably caused by an evolving spot
configuration (see also the discussion in Sect. \ref{sec:discussion}). The
observed out-of-eclipse variability is qualitatively consistent with a large
cool spotted region on the back side of the primary star causing the attenuation
of the secondary eclipse. Also, the secondary eclipse in the ASAS data appears
to be only half the depth that it exhibits in the NMSU-1m data. This is a
further suggestion that the total depth of the secondary eclipse is being
affected by variable starspots.

\subsubsection{Ephemeris}
\label{sec:ephem}

Partial primary and secondary eclipses which were detected in the NMSU-1m data
were used in combination with the ASAS photometry to derive an ephemeris for the
system. We defined the epoch of the system, $T_0$, to be the midpoint of the
most complete primary eclipse. We used the NMSU-1m data obtained on HJD=2454110
which contains the only highly time sampled primary eclipse showing the eclipse
minimum to determine $T_0$. We modelled the shape of this eclipse with a 4th
order polynomial and performed a $\chi^2$ minimization with respect to the model
to derive the best fitting coefficents. We then set $T_0$ to be the midpoint of
the model curve. Next, we fixed $T_0$ and solved for the period of the system
using the nearly complete primary eclipse obtained on HJD=2454110 and the only
other partial primary eclipse (HJD=2454116) in the NMSU-1m data in combination
with the complete ASAS lightcurve. We tested a grid of possible periods,
phase-folding these data on each trial period, and found the period which
minimized the $\chi^2$ of these data with respect to the model. The resulting
ephemeris at the midpoint of the eclipse is given by $$T(V_{\rm min}) = {\rm
HJD}~2454110.8500 \pm 0.002 + 3.8729 E \pm 0.0004 $$ where $E$ corresponds to
the cycle number. The uncertainties were estimated from the curvature of the
$\chi^2$ surface around the minimum, using the standard equation $\sigma =
\sqrt{2 / {\nabla^2 \chi^2}}$. The lack of more than one primary eclipse and the
relatively few ASAS data points with large errors compared to the NMSU-1m data
are the main causes of uncertainties on the ephemeris. We have no additional
constraints on the ephemeris and keep this fixed to the above values during the
eclipsing binary modelling in Sect. \ref{sec:analysis}.

With observations of only one complete primary and secondary eclipse it is
difficult to accurately determine the eccentricity of the system. However, we
estimate this parameter by measuring the time of the midpoint of the complete
secondary eclipse obtained on HJD=2454114 with respect to $T_0$ by modelling the
shape of that eclipse as a 4th order polynomial and minimizing the $\chi^2$. We
find the timing of the secondary eclipse is marginally consistent with a
circular orbit for the binary within the quoted uncertainties on the ephemeris.
Therefore, we adopt an eccentricity $e=0$ for the system.  However, additional
highly time-sampled observations of multiple primary and secondary eclipse
minima are necessary to derive a more accurate ephemeris, and determine whether
there is a small but non-zero eccentricity in the system.
Figure~\ref{fig:folded} shows all the small telescope photometry phase-folded
with the ephemeris defined above.

\subsection{Spectroscopy}

A total of 17 spectra of ASAS~J052821+0338.5 with exposure times of 15--30
minutes were obtained with the Fibre-fed Echelle Spectrograph (FIES) at the 2.5m
Nordic Optical Telescope during January 2--16, 2007. These spectra cover the
wavelength range 4000--7350 {\AA} at a resolution of $R \approx 47\,000$ and
have a typical signal-to-noise of 30--50.

The spectra were reduced with the on-line data reduction software supplied at
the telescope. This package, based on {\sc PYTHON} and {\sc PyRAF} was
especially developed for FIES and performs all conventional steps of echelle
data reduction, including the subtraction of bias frames, modelling and
subtraction of scattered light, flat-fielding, order extraction, normalization
(including fringe-correction) and wavelength calibration. In addition we applied
the heliocentric velocity correction to the wavelength scale and performed
continuum normalization.

Four additional 15-minute high resolution ($R \approx 65\,000$) spectra were
obtained on April 5, 2007 with the UCLES spectrograph and the SEMPOL polarimeter
at the Anglo Australian Telescope. These data were reduced with {\it Echelle
Spectra Reduction: an Interactive Tool} (ESPRIT, Donati et al. \cite{donati97}),
a comprehensive reduction package for spectro-polarimetric observation.

The spectrum of ASAS~J052821+0338.5 is a clear example of a double-lined
spectroscopic binary. Fig. \ref{fig:pmsfeat} shows a sample of the spectrum,
obtained near phase 0.64, with well-separated components. The pre-main-sequence
nature of the object is evident from the H$\alpha$ emission (EW $\approx 1$
{\AA}, secondary only) and strong Li~{\sc i}~6708 absorption. Except for
H$\alpha$, no other emission lines are seen in the spectrum. Some of the deepest
absorption lines in the spectrum, such as the Mg b line complex and Na~{\i}~D
doublet, show weak emission in the line cores. There is no evidence of excess
continuous emission, such as veiling.

\section{Analysis}
\label{sec:analysis}

\subsection{Radial Velocity Curve}
\label{sec:radvel}

To derive the radial velocity curve for the binary system, the 17 FIES spectra
were cross-correlated against the spectrum of \mbox{HD 286264}, a single-lined
K2IV weak-line T Tauri star, on an order-by-order basis using the FXCOR package
in IRAF. The majority of spectra showed two distinct cross-correlation peaks,
one for each component of the binary. Thus, both peaks were fit independently
with a Gaussian profile to measure the velocity and velocity errors of the
individual components. If the two peaks appeared blended, a double Gaussian was
fit to the combined profile. For each of the 17 observations we then determined
a weighted-average radial velocity for each star from all orders without
significant contamination by telluric absorption features. Here we used as
weights the inverse of the variance of the radial velocity measurement in each
order, as reported by FXCOR. The resulting radial velocities were calibrated
using the velocity offset derived by cross-correlating the template spectrum
with a radial velocity standard, HD~109358 (Udry et al. \cite{udry99}). In these
data, we find no evidence for a third component, since the cross-correlation
function showed two distinct peaks. 

To each of the four UCLES spectra, we applied the technique of least-squares
deconvolution (LSD, see Donati et al. \cite{donati97}). This technique, often
used in Doppler Imaging, allows us to construct, with a very high
signal-to-noise, a combined-average line profile from about 1000 absorption
lines in the spectrum. The LSD profile obtained in this way shows two distinct
absorption profiles corresponding to the components of the binary, allowing us
to measure the individual stellar radial velocities.

\begin{figure}
\centering
\includegraphics[angle=0,width=0.95\columnwidth]{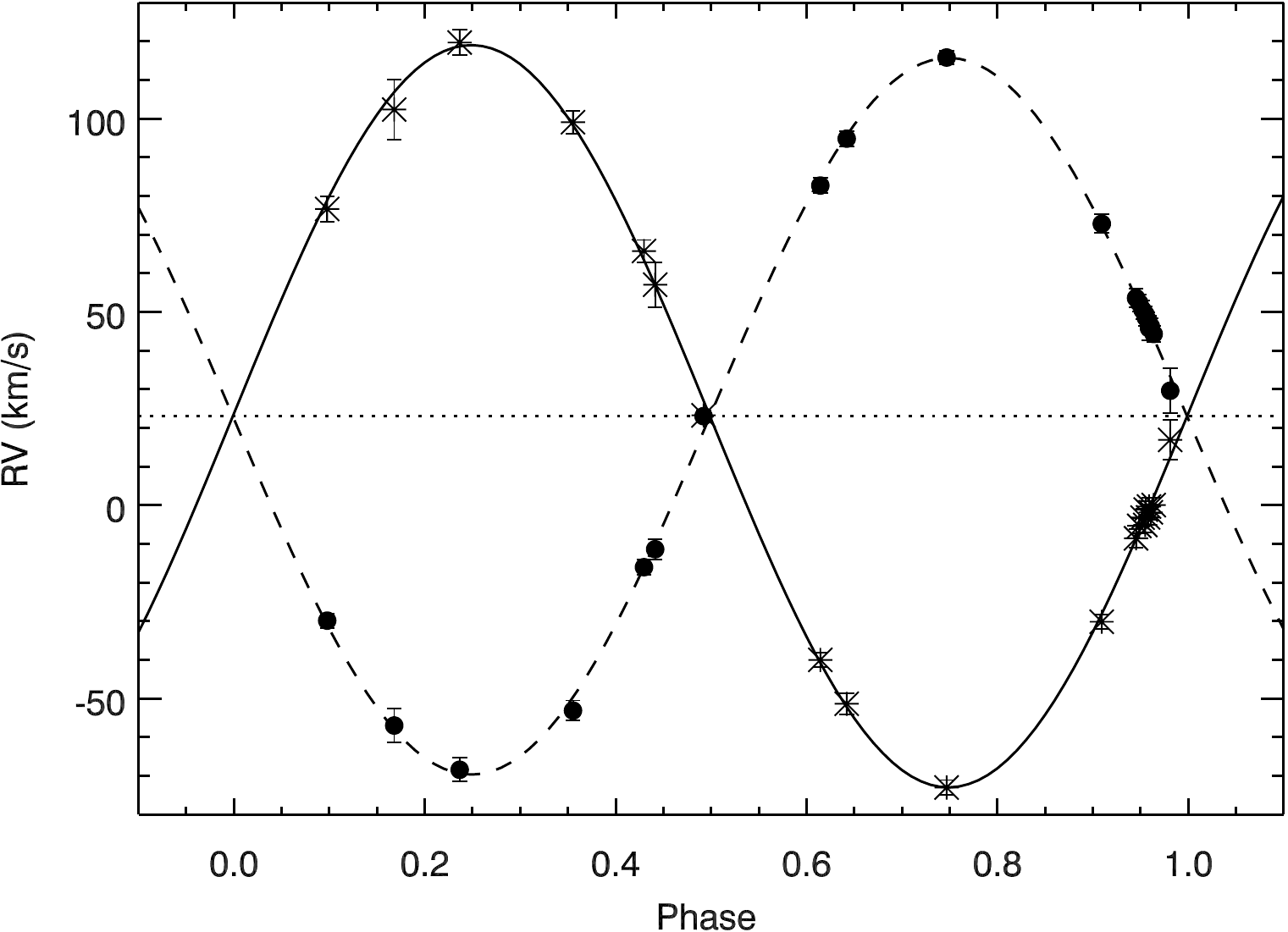}
\caption{Radial velocity measurements of the primary star (filled circles,
dashed line) and secondary star (asteriks, solid line) of ASAS~J052821+0338.5. A
model radial velocity curve for both components is overplotted using the
ephemeris derived in Sect. \ref{sec:ephem}. A binary mass ratio, $M_1/M_2 =
1.03$, provides the best fit to the data. }
\label{fig:rv}
\end{figure}

We determined an initial orbit solution to these radial velocity measurements
using the {\sc binary} code kindly provided by D.~Gudehus, with the orbital
period held fixed at the value determined from the photometric lightcurve (Sect.
\ref{sec:ephem}). To minimize the influence of epochs in which the
cross-correlation functions show highly blended peaks, we excluded two epochs
that occur within 0.05 phase of primary or secondary eclipse. The orbit solution
was then iterated, and the internal error bars adjusted between iterations,
until the final orbit solution yielded a reduced $\chi^2$ of unity. This orbit
solution gives a best-fit eccentricity of $e = 0.02 \pm 0.02$, i.e. formally
consistent with a circular orbit. We then refined this solution by combining the
radial velocity data with the observed lightcurve (see Sect. \ref{sec:lcmodel}).
This solution then yields: a binary mass ratio of $M_1/M_2 = 1.03 \pm 0.005$,
velocity amplitudes of $K_1 = 92.4 \pm 0.33$ and $K_2 = 95.1 \pm 0.56$, and a
system radial velocity of $\gamma = 22.8 \pm 0.3$~km/s. The final radial
velocity curve is shown in Fig.~\ref{fig:rv}.

\subsection{Spectroscopic analysis}
\label{sec:spect}

In order to study each star individually, we numerically reconstructed the
spectra of each of the two components with tomographic separation. This
technique inverts the relation that the observed spectrum is a linear
superposition of the spectra of the individual stars, where the only scaling
factor is the luminosity ratio of the two components, which depends on their
temperatures and radii. Provided that spectra observed at a range of phases are
available, tomographic separation can efficiently reconstruct the individual
stellar spectra, and is now a standard technique for the analysis of double-line
spectrosocopic binaries (see, for example, Bagnuolo \& Gies \cite{bagnuolo91}).
We applied this technique to six spectra, selected on the basis of having the
best signal-to-noise as well as covering a range of orbital phases. Here, we
assumed a luminosity ratio of 2.0 for the two stars, which is consistent with
the ratio of temperatures and radii obtained from lightcurve modelling (see
Sect. \ref{sec:lcmodel}).

From the individual component spectra we then determined the fundamental stellar
parameters using the {\sc IDL}-based spectroscopic analysis package {\sc SME}
(Valenti \& Piskunov \cite{valenti96}). {\sc SME} combines radiative transfer
calculations with multi-dimensional least-squares minimization to determine
which parameters (the effective temperature $T_{\rm eff}$, the gravity $\log g$,
the metallicity $[M/H]$, the projected rotational velocity $v \sin i$, the
systemic radial velocity $v_{\rm rad}$, the microturbulence $v_{\rm mic}$ and
the macroturbulence $v_{\rm mac}$) best describe the observed stellar spectrum.
The grid of model atmospheres used for radiative transfer are taken from Kurucz
(\cite{kurucz93}), and atomic line data were obtained from the VALD database
(Piskunov et al. \cite{piskunov95}; Kupka et al. \cite{kupka99}). The atomic
line data were checked and fine-tuned against a high-quality spectrum of the Sun
(Kurucz et al. \cite{kurucz84}).

The two wavelength regions we considered in our spectrscopic analysis are
5880--5920 {\AA} and 6000--6200 {\AA}. The first region contains the Na {\sc i}
D lines, sensitive to both $T_{\rm eff}$ and $\log g$. The second regions
contains a wealth of non-blended lines of different elements, providing
constraints on $T_{\rm eff}$, $\log g$ and $[M/H]$. We have no independent
leverage on $v_{\rm mic}$ and $v_{\rm mac}$, and therefore assumed $v_{\rm mic}
= 1.6$ km/sec (the average value for K-type PMS stars found by Padgett
\cite{padgett96}) and $v_{\rm mac} = 3.5$ km/sec (for a $T \approx 5000$~K star,
see Gray \cite{gray92}). From this spectroscopic analysis we find for the
primary $T_{\rm eff} = 5103 \pm 100$ K, $\log g = 4.0 \pm 0.1$ and $[M/H] = -0.2
\pm 0.2$, and for the secondary $T_{\rm eff} = 4705 \pm 100$ K, $\log g = 3.9
\pm 0.1$ and $[M/H] = -0.1 \pm 0.2$. These parameters correspond to spectral
types of K1 for the primary and K3 for the secondary (Cohen \& Kuhi
\cite{cohen79}).

The projected rotational velocities of the two components were determined by
analysing the LSD line profiles (see Sect. \ref{sec:radvel}) of each spectrum.
During this process, we took into account line broadening from other sources,
such as macroturbulence, smearing due to the orbital motion of the binary during
the exposure and the instrumental profile. We find an average $v \sin i$ of
$24.5 \pm 0.8$ km/sec and $24.5 \pm 0.7$ km/s for, respectively, the primary and
secondary. These values suggest the stars rotate synchronously (see Sect.
\ref{sec:discussion}).

In addition to the fundamental stellar parameters, we determined the abundance
of lithium from the prominent resonance line at 6708 {\AA} (EW${}_{\rm prim}
\approx 307$\,m{\AA}, EW${}_{\rm sec} \approx 394$\,m{\AA}). After applying
corrections of $-0.29$ (primary) and $-0.21$ (secondary) for NLTE effects
(Carlsson et al. \cite{carlsson94}), we obtained $\log n({\rm Li}) \equiv (Li/H)
+ 12 = 3.10 \pm 0.2$ for the primary, and $\log n({\rm Li}) = 3.35 \pm 0.2$ for
the secondary. These values are consistent with the commonly adopted value of
3.2--3.3 for the primordeal lithium abundance.

\begin{figure}
\centering
\includegraphics[angle=90,width=0.95\columnwidth]{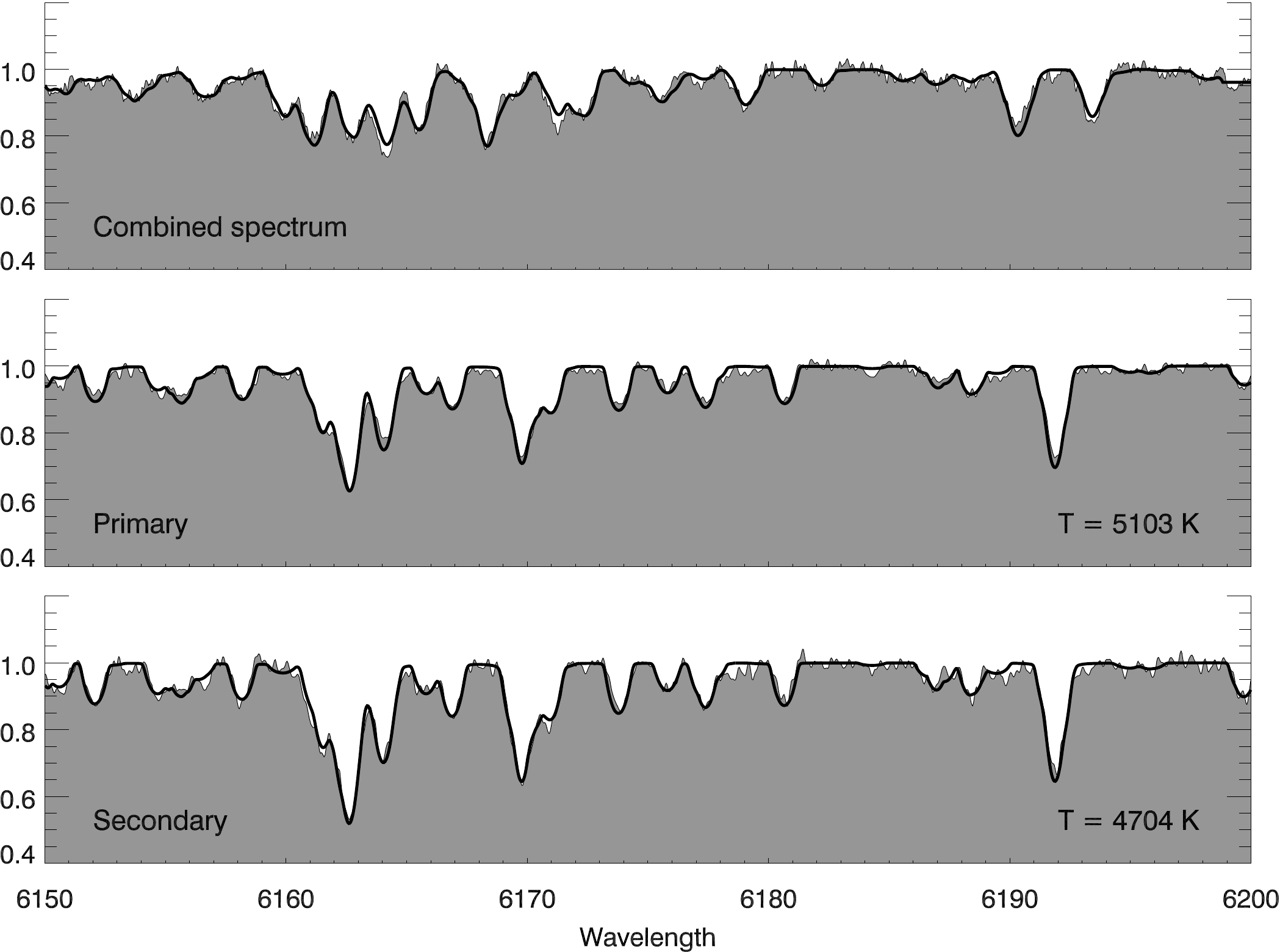}
\caption{The above panels show 50 {\AA} of the observed spectrum of
ASAS~J052821+0338.5 (upper panel), as well as the tomographically reconstructed
spectra of the primary (middle panel) and the secondary component (lower panel),
both in their respective stellar rest frames. The thick solid lines in each
panel are synthetic spectra calculated with the stellar parameters from
Table~\ref{tab:props}.}
\label{fig:pmssynth}
\end{figure}

Examples of tomographically separated spectra, as well as a comparison between
the observed and synthetic spectra for each star are shown in
Fig.~\ref{fig:pmssynth}. A full compilation of the stellar and atmospheric
parameters is presented in Table \ref{tab:props}.

\subsection{Lightcurve modelling}
\label{sec:lcmodel}

Together with determining the stellar and orbital parameters from spectroscopy
we fit the $V$-band lightcurve and radial velocity data using the most recent
version of the eclipsing binary lightcurve modeling algorithm of Wilson \&
Devinney (\cite{wd71}, with updates), as implemented in the {\sc phoebe} code of
Pr{\v s}a  \&  Zwitter (\cite{prsa05}). The code determines the surface gravity
and effective temperature of each star based on the calculated gravitational
potential, assuming full Roche geometry according to the formalism of Kopal
(e.g. Wilson \cite{wilson79}), and the gravity brightening coefficients from
Claret (\cite{claret00}). It includes theoretical Kurucz model atmospheres to
determine intensities over the stellar disks. We adopted a linear limb darkening
law using coefficients extrapolated down to 2700 K from Van Hamme
(\cite{vanhamme93}). We allowed the code to calculate reflection effects,
adopting a bolometric albedo of 0.5, typical for a fully convective stellar
envelope. Since the out-of-eclipse variations in the observed lightcurve suggest
the presence of starspots which are changing over time, we include in the
modelling only $V$-band photometry obtained with the NMSU-1m and the JGT-1m
between January 9--February 10, 2007.

To begin, we first applied a traditional rectification procedure in  which we
fit the out-of-eclipse variations with a smooth polynomial function. The aim of
this rectification procedure is to ``remove" the spot signal from the lightcurve
and to thereby allow the lightcurve model to achieve goodness of fit without
introducing additional spot-fitting parameters that are poorly constrained at
present.

Next, in order to maintain control of the lightcurve solution and its many free
parameters, we performed this fitting in stages. In the initial stage we held
the orbital parameters fixed at the values determined from the double-lined
spectroscopic orbit solution (see Sect. \ref{sec:radvel}), and fixed the
effective temperature of the primary component to the value determined from our
spectral analysis (5103~K; see Sect. \ref{sec:spect}). This allowed us to obtain
initial estimates of the component temperatures and radii, and the system
inclination.

With initial values for the temperatures, radii, and  inclination so determined,
we then iteratively improved the solution by first allowing the eccentricity and
argument of periastron to be fit, and then performing a final fit in which all
of the orbital and component parameters of the system were fit freely (except
for the ephemeris, see Sect. \ref{sec:ephem}). The eccentricity we recovered for
the system, $e = 0.007 \pm 0.006$, is indistinguishable from a circular orbit,
and we could therefore not constrain the periastron angle. Also, if we assume a
truely circular orbit ($e = 0$) we find no measurable change in any of the
orbital parameters (within the limits of the error bars). We have therefore
chosen to list this system as circular, and report no periastron angle or time
for periastron passage. The final parameters that result from this simultaneous
fit to the radial velocities and rectified lightcurve are presented in Table
\ref{tab:props}.

To explore the influence of, and origin of the out-of-eclipse variability, and
to better characterize systematic uncertainties in the stellar parameters
arising from this variability, we have also modelled the un-rectified lightcurve
by introducing multiple spots into the fitting analysis. The number of spots,
their placement on the stellar surfaces, their temperatures and sizes, were
initially determined manually in order to approximate the strong out-of-eclipse
variability, then the code was permitted to adjust the spot parameters along
with all of the other system and component parameters. We wish to emphasize that
the adopted spot solution is neither definitive nor unique. Degenerate starspot
configurations are consistent with the observed out-of-eclipse photometry, and
it is not possible to break the degeneracy with the current single-band
lightcurve. Therefore we present only one possible starspot configuration which
well mimics the observed out-of-eclipse variations. Despite the degeneracy of
the starspot solutions, the lightcurve cannot be adequately fit without the
presence of a large ($\sim 25^{\circ}$ radius) cool spot on the back of the
primary star. We also include a small hot spot near the back of the secondary
star (i.e.\ facing away from the primary) in order to reproduce the small
``bump" just prior to the primary eclipse. In Table \ref{tab:props} we summarize
the stellar parameters resulting from simultaneously fitting the radial velocity
and un-rectified lightcurve data, for ease of comparison with the parameters
resulting from fitting the rectified lightcurve. The different treatment of
modelling the distribution of starspots did no result in any change in the
orbital parameters, and these are therefore not repeated in Table
\ref{tab:props}.

The stellar parameters resulting from the spot modelling procedure are overall
in good agreement with those resulting from the rectification procedure. The
formal errors on the stellar parameters derived from the above analysis are
given in Table~\ref{tab:props}, but these do not include systematic errors
inherent to the modelling procedures. Still, by comparing the results from the
two methods (spotted vs. rectified) we find that, despite the obviously strong
effect of spots on the lightcurve, a detailed treatment of the spots is
evidently not critical for most of the stellar parameters of interest. The
notable exceptions are the temperature and radius of the secondary. These
parameters depend on the eclipse depths, which are affected by the manner in
which the out-of-eclipse variations are modelled.

\begin{figure*}
\includegraphics[angle=0,width=\textwidth]{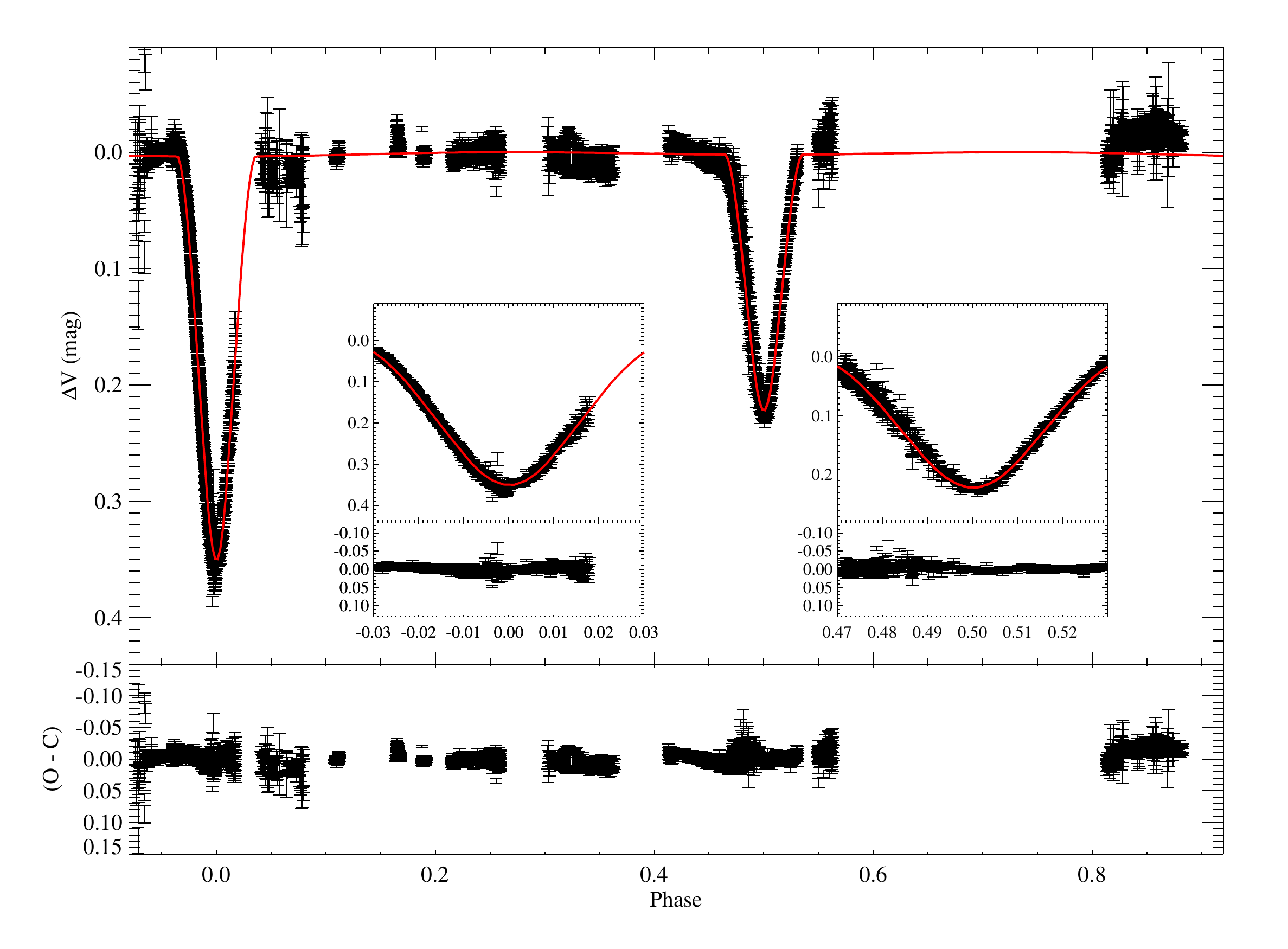}
\caption{The top panel shows the phase-folded and rectified differential
$V$-band lightcurve obtained with the NMSU-1m and JGT-1m telescopes. The
lightcurve has a precision of $0.005$--$0.01^{\rm m}$ and a sampling of $<
1$~min. The data are phase-folded with a period, $P =3.8729$~days, and epoch,
$T_0 = {\rm HJD}~2454110.8500$, as well as rectified by fitting a smooth
polynomial to the out-of-eclipse variations. The smooth grey curve shows the
calculated binary lightcurve that has the best agreement with the data. The
residuals of this curve with the data are shown in the bottom panel.}
\label{fig:lc}
\end{figure*}

The temperature of the secondary is determined relative to the temperature of
the primary (which we hold fixed at 5103~K) by virtue of the temperature ratio,
which is principally determined from the ratio of eclipse depths. Between the
two methods of analysis, we find a difference of $\sim 150$ K in the inferred
temperature of the secondary, a difference that is significantly larger than the
formal uncertainty of $\sim 25$ K in the lightcurve model (Table
\ref{tab:props}). The temperature resulting from the rectified analysis is more
consistent with the value from spectroscopic analysis of the tomographically
separated spectra (Sect. \ref{sec:spect}), thus in what follows we adopt this
solution as the preferred one. This lightcurve model is overplotted on the
phase-folded data in Fig.~\ref{fig:lc}.

The radii obtained from the two light-curve solutions also show larger
differences than expected from the formal errors. Even so, the  sum of the radii
is consistent in both solutions to within $\sim 2$\%. Due to the circular orbit,
the primary and secondary eclipses are of equal durations and thus it is not
possible to determine the individual radii independently from a single-band
light curve. While the {\it sum} of the radii is very well constrained by the
total duration of the two eclipses, almost any combination of component radii
that satisfy the sum will yield a good fit to the light curve. Of course, the
different component radii will in general result in very different flux ratios
(since the temperature ratio is well constrained). Thus, we caution that
additional observational constraints (e.g. flux ratios at multiple wavelengths)
will be needed to firmly establish the individual radii accurately. The
spectroscopically derived temperatures cannot by themselves constrain these flux
ratios, since tomographic separation requires us to assume a luminosity ratio
(see Sect. \ref{sec:spect}). We have therefore iteratively worked towards a
solution for the temperatures and radii that is consistent with both the
spectroscopic and the photometric models.

\begin{table}

\caption{Properties of ASAS~J052821+0338.5. For each component we list the
atmospheric parameters: effective temperature ($T_{\rm eff}$), surface gravity
($\log g$), rotational velocity ($v \sin i$), metallicity ($[M/H]$) and lithium
abundance ($\log n({\rm Li})$). We also present the systems ephemeris: the
ephemeris reference epoch ($T_0$) period ($P$), eccentricity ($e$), systemic
radial velocity ($\gamma$), inclination ($i$) and projected semi-major axis ($a
\sin i$). Since the system's eccentricity is indistinguishable from truly
circular, we choose $e=0$ and list no periastron angle or time for periastron
passage. Finally, we list the parameters of our solutions from lightcurve
modelling with and without spots: the velocity amplitude ($K$), stellar mass
($M$), radius ($R$), luminosity ($L$) and effective temperature ($T_{\rm eff}$)
}
\label{tab:props}
\centering
\begin{tabular}{l r r l}
\hline\hline                  
Parameter & Primary & Secondary & Unit\\
\hline                        
\multicolumn{4}{c}{\it Stellar parameters from spectral analysis}\\
$T_{\rm eff}$ & $5103  \pm 100 $  & $4705   \pm 100 $ & K             \\
$\log g$      & $ 4.0  \pm 0.1 $  & $ 3.9   \pm 0.1 $ &                 \\
$v \sin i$    & $24.5  \pm 0.8 $  & $24.5   \pm 0.7 $ & km s${}^{-1}$ \\
$[M/H]$       & $-0.2  \pm 0.2 $  & $-0.1   \pm 0.2 $ &                 \\
$\log n({\rm Li})$ & $3.10  \pm 0.2 $  & $3.35   \pm 0.2 $ &            \\
 & & &     \\
\multicolumn{4}{c}{\it Parameters from rectified lightcurve modelling}\\
$T_0$         & \multicolumn{2}{c}{${\rm HJD}~2454110.8500 \pm 0.005$} &      \\
$P$           & \multicolumn{2}{c}{$3.8729 \pm 0.0002$} & d          \\
$e$           & \multicolumn{2}{c}{$0$} &                 \\
$\gamma$      & \multicolumn{2}{c}{$22.8   \pm 0.3$} & km s${}^{-1}$ \\
$i$           & \multicolumn{2}{c}{$83.7   \pm 1.0$} & deg           \\
$a \sin i$    & \multicolumn{2}{c}{$0.0668  \pm 0.0002$} & AU        \\
$K$           & $92.41 \pm 0.33$  & $95.08 \pm 0.56$ & km s${}^{-1}$ \\
 & & &     \\
$M$           & $1.375 \pm 0.011$      & $1.329 \pm 0.008$ & $M_{\sun}$ \\
$R$           & $1.83  \pm 0.01$       & $1.73  \pm 0.01$ & $R_{\sun}$  \\
$L$           & $2.05   \pm 0.16$      & $1.38  \pm 0.11$ & $L_{\sun}$  \\
$T_{\rm eff}$ & $5103^{*}$             & $4751  \pm 26$   & K           \\
$\log g$      & $4.05  \pm 0.01$       & $4.08  \pm 0.01$ &             \\
 & & &     \\
\multicolumn{4}{c}{\it Parameters from EB modelling with spots}\\
$M$           & $1.387 \pm 0.017$     & $1.331 \pm 0.011$ & $M_{\sun}$  \\
$R$           & $1.84  \pm 0.01$      & $1.78  \pm 0.01$  & $R_{\sun}$  \\
$L$           & $2.06  \pm 0.16$      & $1.28  \pm 0.10$  & $L_{\sun}$  \\
$T_{\rm eff}$ & $5103^{*}$	       & $4599  \pm 23$    & K  	 \\
$\log g$      & $4.05  \pm 0.02$      & $4.06   \pm 0.01$ &             \\
\hline\hspace{2mm}\\
\multicolumn{4}{l}{${}^{*}$ Assumed parameter for EB modelling}\\
\end{tabular}
\end{table}

\subsection{Distance, physical location and kinematics}
\label{sec:phys}

We estimated the distance to ASAS~J052821+0338.5 by comparing distance moduli
from the available broad-band magnitudes. For this analysis, we have assumed
relatively large uncertainties on the observed magnitudes ($0.15^m$), because we
have no possibility of knowing whether these are truly out-of-eclipse values.
Using the components' surface temperatures of the preferred solution from
lightcurve modelling (5103~K and 4751~K) and the standard color indices
tabulated by Kenyon \& Hartmann (\cite{kenyon95}), we find that the distance
moduli of the available colors converge on a distance of $280 \pm 30$ pc and an
extinction of $A_V = 0.6 \pm 0.3^{\rm m}$. The distance is mostly constrained by
the (almost) extinction-insensitive $J$, $H$ and $K$ magnitudes, while the
extinction depends mostly on the $B$ and $V$ magnitudes. The quoted error
margins are dominated by the uncertainties of the observed magnitudes.

In addition to the above, we compared the observed broad-band magnitudes of
ASAS~J052821+0338.5 with model stellar atmosphere spectra from Castelli \&
Kurucz (\cite{castelli03}), using the SED fitting routines described in
Robitaille et al. (\cite{robitaille07}; modified to incorporate stellar
atmosphere models). Assuming a temperature of 5000 K, which is the
luminosity-weighted average temperature of the binary, we find the broad-band
magnitudes to be compatible with an extinction of $A_V \approx 0.65^{\rm m}$. We
also measured the equivalent width of the narrow interstellar absorption
components in the Na {\sc i} D1 line. We find two absorption components with a
combined EW of about 200 m{\AA}, which translates to $A_V \approx 0.25 \pm
0.1^{\rm m}$ (Munari \& Zwitter \cite{munari97}). The interstellar Na~{\sc i}~D
absorption is partly filled-in by telluric emission, and therefore this value
should be considered a lower limit to $A_V$. We notice that the strongest
absorption component (with an EW of $\sim 150$ m{\AA}) has a central velocity
equal to the systemic velocity of the system, suggesting that a large part of
the extinction originates from regions close to the binary.

Following the definitions of Warren \& Hesser (\cite{warren77}),
ASAS~J052821+0338.5 is located in the direction of the widely spread Orion OB1a
subassociation. The system is about 3--4 degrees away from the much more compact
Orion OB1b subassociation and about 2 degrees from the 25 Ori subassociation, a
compact and kinematically distinct subgroup of Orion OB1a recently identified by
Brice{\~n}o et al. (\cite{briceno07}). Orion OB1a is located at a distance of
$330 \pm 15$ pc, while Orion OB1b is further away, at a distance $440 \pm 20$ pc
(Brown et al. \cite{brown94}; de Zeeuw et al. \cite{dezeeuw99}; Hern{\'a}ndez et
al. \cite{hernandez05}). There is little difference in the kinematic
distribution of Orion OB1a and Orion OB1b. The average radial velocities of
Orion OB1a and Orion OB1b, $23.8 \pm 0.7$ km/sec and $23.1 \pm 1.4$ km/sec
respectively (Morrell \& Levato \cite{morrell91}), and also their tangential
velocity distributions are similar ($\mu_\alpha = 0.44 \pm 0.25$ mas/yr,
$\mu_\delta = -0.65 \pm 0.25$ mas/yr, Brown et al. \cite{brown98}; de Zeeuw et
al. \cite{dezeeuw99}). Orion OB1a is commonly assumed to be $11.4 \pm 1.9$ Myr
old, significantly older than the $1.7 \pm 1.1$ Myr old Orion OB1b (Brown et al.
\cite{brown94}).

The systemic radial velocity we determined for ASAS~J052821+0338.5 is $22.8 \pm
0.3$ km/sec, and the tangential velocity from the NOMAD catalog is $\mu_\alpha =
-1.6 \pm 2.8$ mas/yr, and $\mu_\delta = -4.0 \pm 4.0$ mas/yr. These kinematics
combined with the estimated distance and physical location suggest that
ASAS~J052821+0338.5 is most probably a member of the $\sim 11$~Myr old Orion
OB1a subassociation.

\subsection{Evolutionary status}

Several characteristics of the two binary components, including their large
radii, their cool temperatures, and their primordial lithium abundances, as well
as the proximity to the Orion OB1a subassociation strongly indicate that
ASAS~J052821+0338.5 is a young, pre-main sequence eclipsing binary.

\begin{figure}
\centering
\includegraphics[width=0.95\columnwidth]{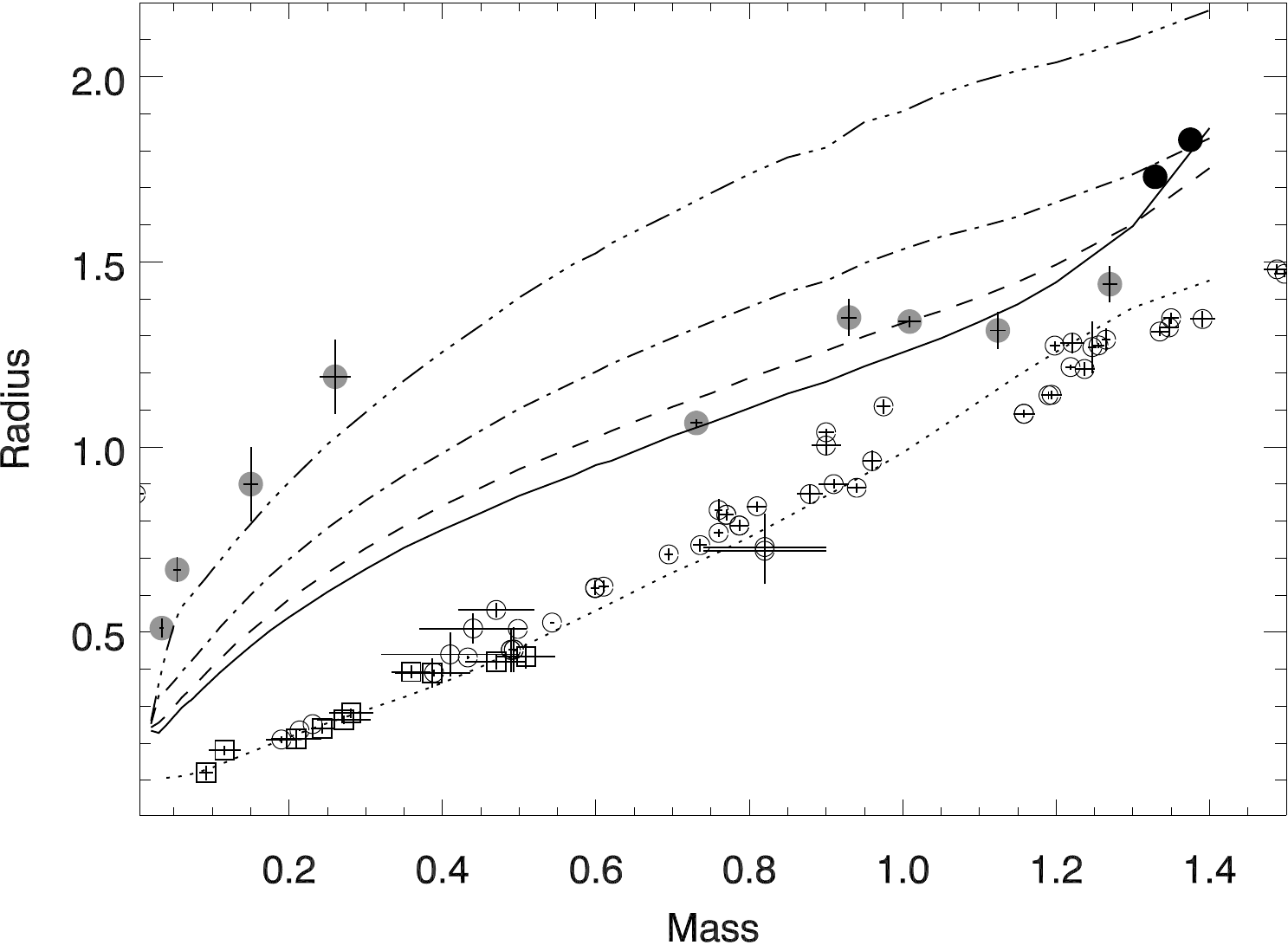}
\caption{Mass versus radius for known main sequence (open circles) and pre-main
sequence (grey solid circles) eclipsing binaries. The components of
ASAS~J052821+0338.5 are shown as black solid circles. Isochrones from Baraffe et
al. (\cite{baraffe98}) with ages of 3~Myr, 6~Myr, 10~Myr, 13~Myr, and 300~Myr
are overplotted. }
\label{fig:mrplot}
\vspace{5mm}
\end{figure}

In Figure~\ref{fig:mrplot}, the structural properties of the binary are shown in
relation to other PMS and main-sequence eclipsing binaries, as well as
theoretical stellar evolution models from Baraffe et al. (\cite{baraffe98}).
Both components have larger radii than main-sequence objects of the same mass,
as expected for stars that are still evolving onto the main sequence. However,
their large observed radii do not alone imply a pre-main-sequence evolutionary
state. A typical, post-main-sequence, $1.4\,{\rm M}_{\sun}$ star with solar
metallicity reaches a radius similar to that of the primary component of
ASAS~J052821+0338.5 at an age of $\sim 2$~Gyr. Yet, at about 6000\,K, such a
star is considerably hotter than either component of ASAS~J052821+0338.5. A
$1.4\,{\rm M}_{\sun}$ post-main-sequence star will cool to a temperature
comparable with ASAS~J052821+0338.5 ($\sim 5000$\,K) as it evolves through the
red giant phase, but this necessarily coincides with a further increase in
radius to a value much larger than what is observed in the binary. Thus, the
cool temperatures of the binary components combined with their $\sim 1.4\,{\rm
M}_{\sun}$ masses and $< 2 {\rm R}_{\sun}$ radii suggest ASAS~J052821+0338.5 is
a pre-main sequence object.

This is exemplified in the modified Hertzprung-Russel (HR) diagram shown in
Fig.~\ref{fig:hrplot}. In this figure, stellar radius is a proxy for luminosity,
as it is a direct measurement from the eclipsing binary analysis, and it relates
to luminosity through the temperature. Evolutionary tracks and isochrones for
the Baraffe et al. (\cite{baraffe98}) models are shown in addition to the
position of ASAS~J052821+0338.5 and the other known PMS eclipsing binaries. We
interpolated the evolutionary tracks at the masses of the primary (black solid
line) and secondary (black dashed line) components. ASAS~J052821+0338.5 is
clearly positioned in the PMS phase of evolution in this parameter space when
compared to empirical data and to this set of models.

A further indication of the PMS evolutionary state of ASAS~J052821+0338.5 is the
detection of primordial lithium in both components. The rate of lithium
depletion in late-type stars depends on the depth of the convection zone (and
thus, on temperature), and the primary is therefore expected to have the 
slowest lithium depletion rate. The primary, with an effective temperature of 
5100\,K, will deplete lithium well below the primordial level after about
150--200 Myr (Sestito \& Randich \cite{sestito05}). In conclusion, only very
young stars can have bloated radii, cool temperatures, and primordial lithium.

Finally, although determining an absolute age of ASAS~J052821+0338.5 would
involve a comparison with largely uncalibrated stellar evolutionary models, we
do note that the fundamental parameters of ASAS~J052821+0338.5 are consistent
with the $10$--$13$ Myr solar-metallicity isochrones of Baraffe et al.
(\cite{baraffe98}). This is in agreement with the age of the Orion OB1a
subassociation ($11.4 \pm 1.9$ Myr, see Sect.~\ref{sec:phys}).

\section{Discussion}
\label{sec:discussion}

\begin{figure}
\centering
\includegraphics[width=0.95\columnwidth]{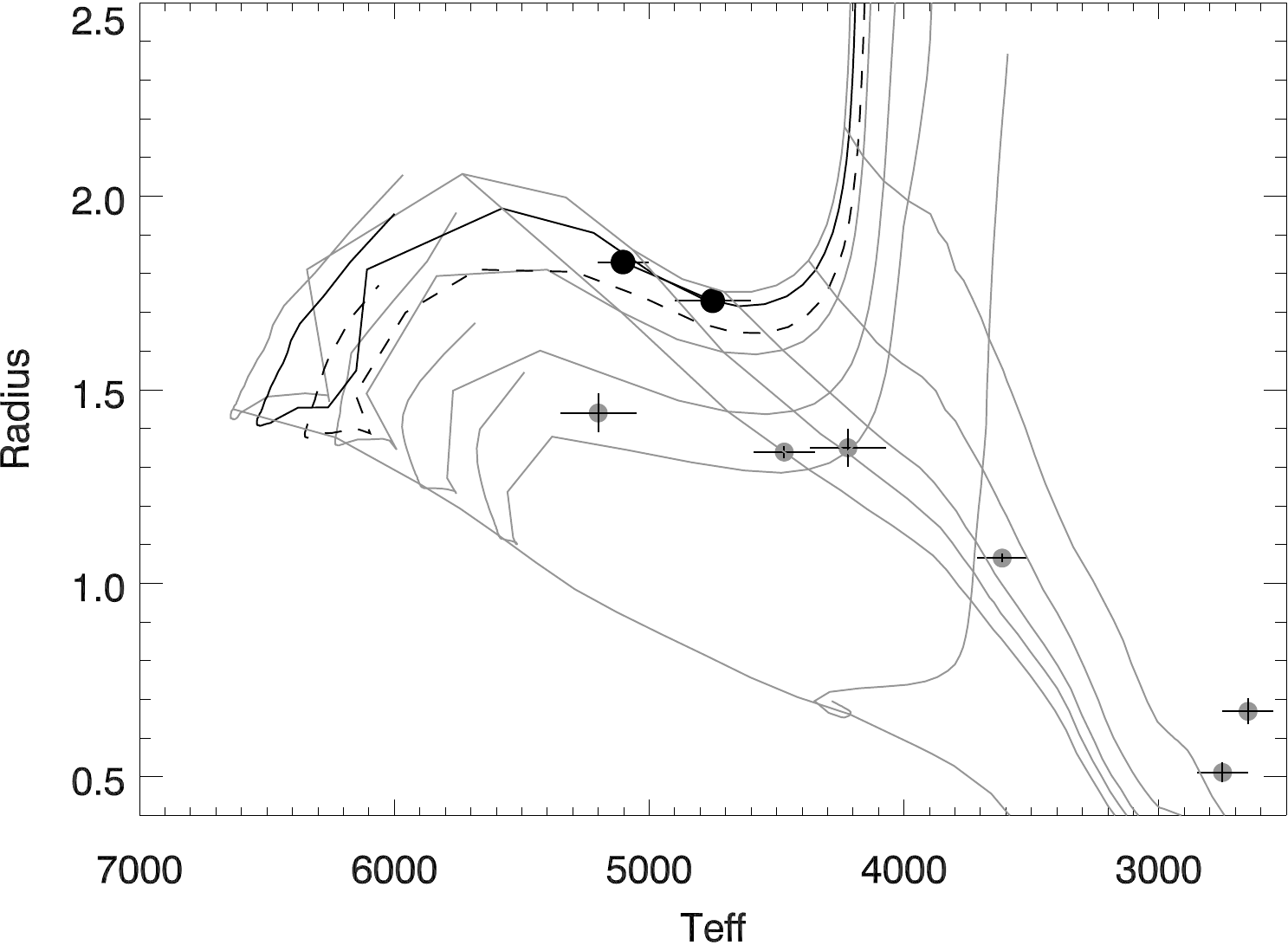}
\caption{The modified Hertzprung-Russel diagram showing radius versus effective
temperature for the components of ASAS~J052821+0338.5 (black solid circles) and
other PMS EBs in Orion (grey solid circles). Evolutionary tracks (1.4, 1.3, 1.2,
1.1, 0.7 M$_{\sun}$) and isochrones (3, 6, 10, 13, 16, 300 Myr) from Baraffe et
al. \cite{baraffe98} are overplotted as grey lines. The black solid line shows
the evolutionary track interpolated at the mass of the primary star
(1.375~M$_{\sun}$). The black dashed line shows the evolutionary track
interpolated at the mass of the secondary star (1.329~M$_{\sun}$). }
\label{fig:hrplot}
\end{figure}

The photometric and spectroscopic observations we obtained of
ASAS~J052821+0338.5 identify this system as a new pre-main-sequence eclipsing
binary with a period of $3.87$ days. The masses and radii of the components
place the system in an as yet unsampled region of the mass-radius and HR
diagrams, and thereby provide a new pair of reference points for PMS stellar
evolutionary models. Both components of the system are more massive than all
other members of known PMS eclipsing systems, except for TY~CrA~B, the 1.64
${\rm M}_{\sun}$ and $\sim 3$ Myr old companion to a main-sequence B-type
primary (Casey et al. \cite{casey98}).

The physical location and kinematic properties of ASAS~J052821+0338.5 suggest
this system to be a member of the $\sim 11.4 \pm 1.9$~Myr Orion OB1a
subassociation. This age is also compatible with the position of the two
components in the mass-radius diagram shown in Fig.~\ref{fig:mrplot}, where both
stars fall close to the 13~Myr isochrone of the Baraffe et al.
(\cite{baraffe98}) evolutionary models. At this age, the two 1.35 ${\rm
M}_{\sun}$ stars have just left the fully-convective Hayashi stage. The
temperature of the two components will then increase at almost constant
luminosity, first slightly growing in radius before their final contraction
towards the main sequence, settling at a final temperature of $\sim 6250$\,K
(for a more detailed description of pre-main-sequence evolution, see for example
Palla \& Stahler \cite{palla93}). We also find a that the locations of the stars
in the HR diagram agree quite well with the predicted 13 Myr evolutionary tracks
of Baraffe et al. (\cite{baraffe98}, see Fig.~\ref{fig:hrplot}). Although there
appear to be small differences between the observed radii and the predicted
values, this does not indicate any inconsistency since the only well-constrained
parameter is the sum of the radii (see Sect. \ref{sec:lcmodel}), and full
consistency of both stars with the predicted tracks requires only small
antagonal adjustements (of about 3\%) to the stellar radii. Therefore, a
critical next step to further improve these parameters will be a multi-band
lightcurve analysis with a more detailed spot treatment.

Most of the known PMS eclipsing binaries are located in Orion.
ASAS~J052821+0338.5 and RXJ~0529.4+0041A are thought to be located in the Orion
OB1a subassociation (Covino et al. \cite{covino00}, Covino et al.
\cite{covino04}). V1174~Ori is considered to be a member of Orion OB1c (Stassun
et al. \cite{stassun04}) and 2MASS~J05352184-20130546085, JW 380 and Par 1802
are found in the Orion Nebula Cluster (Stassun et al. \cite{stassun06}; Irwin et
al. \cite{irwin07b}; Cargile et al. \cite{cargile07}). These OB subassociations
in Orion form an evolutionary sequence, with ages of $\sim$\,1--12 Myr. This age
range covers a key period in star formation when stars experience rapid
evolution in both radius and temperature. The components of the PMS eclipsing
binaries in Orion have masses between 0.03 and 1.35~${\rm M}_{\sun}$ and have
well-determined radii and temperatures. As such they form an easily accessible
and homogenous sample of late-type PMS stars that can be used to perform
relative calibrations of PMS stellar evolutionary models.

As described in Sect. \ref{sec:phot}, we detected considerable ($\sim~0.15^{\rm
m}$) out-of-eclipse variations, occuring around the same orbital phase
throughout our four-month campaign of photometric observations, not unlike the
`photometric wave' that is characteristic of the chromospherically active RS~CVn
binaries (see the review by Rodon{\'o} \cite{rodono92}). Like in an RS~CVn
binary, the components of ASAS~J052821+0338.5 have relatively short rotational
periods and extensive convective envelopes, which help to drive a stellar dynamo
and enhance magnetic activity. The photometric variations may then be explained
by a large cool spot (or region of spots) located on the back of the primary,
rotating in and out of view in phase with the binary period. However, this
configuration requires the primary to rotate synchronously with the binary
orbit. According to the theory of Zahn \& Bouchet (\cite{zahn89}), tidal
breaking during the Hayashi phase, when stars are fully convective, is a very
efficient mechanism for inducing orbital synchronization and circularization in
short-period late-type binaries. Once the convective envelope starts decreasing,
tidal breaking loses its efficiency, and further stellar contraction may lead to
(temporary) departures from synchronous rotation. ASAS~J052821+0338.5 seems to
follow this pattern. The orbit is almost perfectly circularized, and the
observed rotational velocities of the primary and secondary are $24.5 \pm 0.8$
km/sec and $24.5 \pm 0.7$ km/s, respectively. Since the expected synchronous
rotational velocities are $24.0$ km/sec and $23.4$ km/sec, the primary is most
likely in synchronous rotation with the binary orbit, while the smaller
secondary may rotate slightly super-synchronously.

While an extended cool spot on the back of the primary can fully explain the
observed out-of-eclipse photometric variations, we did not detect any signatures
of a cool spot in our current high-resolution spectra, such as changes in the
observed line-ratios or the appearance of molecular features. However, the
brightness contrast between the spotted and the unspotted areas is large, and
since the signal-to-noise of our spectra is moderate it is unclear whether such
an effect would be detectable. An alternative explanation to the out-of-eclipse
variability might be periodic eclipses of the system by a warped circumbinary
disk, or by circumstellar material in co-rotation with the binary. Such
concentrations of circumstellar matter localized around the stable co-rotating
Lagrangian points (L2 and L3) have been detected in the close PMS binary V4046
Sgr (Stempels \& Gahm \cite{stempels04}). We are currently pursuing multi-band
photometry of the system in order to determine the nature of the out-of-eclipse
variations, as well as to further improve the orbital and stellar parameters of
ASAS~J052821+0338.5, in particular the temperatures and radii of the components.

\begin{acknowledgements}
We thank Jonathan Irwin for obtaining photometry with the Cambridge 0.5m
3-Mirror Telescope. HCS acknowledges support from the Swedish Research Council.
KGS gratefully acknowledges support from National Science Foundation Career
grant AST-0349075. The Nordic Optical Telescope is operated on the island of La
Palma jointly by Denmark, Finland, Iceland, Norway, and Sweden, in the Spanish
Observatorio del Roque de los Muchachos of the Instituto de Astrofisica de
Canarias.
\end{acknowledgements}

\end{document}